\documentclass[aps,pre,twocolumn,superscriptaddress]{revtex4-1}
\usepackage{graphics,amssymb,amsmath,epsfig,color}

\begin{document}


\title{Nash equilibrium and evolutionary dynamics in semifinalists' dilemma}


\author{Seung Ki Baek}
\email[]{seungki@pknu.ac.kr}
\affiliation{Department of Physics, Pukyong National University, Busan 608-737,
Korea}

\author{Seung-Woo Son}
\email[]{sonswoo@hanyang.ac.kr}
\affiliation{Department of Applied Physics, Hanyang University, Ansan 426-791,
Korea}

\author{Hyeong-Chai Jeong}
\email[]{hcj@sejong.ac.kr}
\affiliation{Department of Physics, Sejong University, Seoul 143-747, Korea}

\date{\today}

\begin{abstract} 
We consider a tournament among four equally strong semifinalists.
The players have to decide how much stamina to use in the semifinals,
provided that the rest is available in the final and the third-place
playoff. We investigate optimal strategies for allocating stamina to the
successive matches when players' prizes (payoffs) are given according to the
tournament results.
From the basic assumption that the probability to win a match follows a
nondecreasing function of stamina difference,
we present symmetric
Nash equilibria for general payoff structures.
We find three different phases of the Nash equilibria in the payoff space.
First, when the champion wins a much bigger payoff than the others,
any pure strategy can constitute
a Nash equilibrium as long as all four players adopt it in common.
Second, when the first two places are much more valuable than the other two,
the only Nash equilibrium is such that everyone uses
a pure strategy investing all stamina in the semifinal.
Third, when the payoff for last place is much smaller than the others,
a Nash equilibrium is formed when every player adopts
a mixed strategy of using all or none of its stamina in the semifinals.
In a limiting case that only last place pays the penalty,
this mixed-strategy profile can be proved to be a unique symmetric Nash
equilibrium,
at least when the winning probability follows a Heaviside step function.
Moreover,
by using this Heaviside step function, we study the tournament by using
evolutionary
replicator dynamics to obtain analytic solutions, which reproduces the
corresponding Nash equilibria on the population level and gives information
on dynamic aspects.
\end{abstract}



\maketitle

\section{Introduction}
 \label{sec:int}
During the 1938 FIFA World Cup in France, Adhemar Pimenta, the coach
of the Brazilian national team, was facing a dilemma: Spearheaded by
Le\^onidas da Silva, the Brazilians had defeated the Polish in extra
time and consecutively eliminated Czechoslovakia in a replay after the ``Battle
of Bordeaux.'' Pimenta was becoming worried about the fatigue
accumulation of his team members.  The next opponent in the semifinal was Italy,
the reigning world champion at that time. Worse was that the final with
Hungary would not be easier.  After changing the starting lineup eight times,
the coach decided to play the semifinal against Italy without Le\^onidas.
Then Brazil lost to Italy.

One might say retrospectively that the coach was imprudent. However,
it is not a trivial question what would have been a better choice in that
situation,
especially for the person directly involved.  Let us look back at their next
encounter during the 1970 World Cup in Mexico. This time, Italy advanced to the
final after the ``Game of the Century'' against Beckenbauer's Germany.
However, the energy exhausted
in the semifinal turned out to be such a great loss that the Azzurri was
utterly defeated by the Brazilians in the final. So they had to watch
helplessly as Brazil took permanent ownership of the Jules Rimet Trophy.

This kind of dilemma is by no means rare in tournament competitions.
It might have originated from the unfairness of sports tournament compared to sports leagues, which is traded off against the efficiency of tournament competitions~\cite{BenNaim2007a,BenNaim2007b,BenNaim2007c,BenNaim2013}.
How to distribute stamina over a series of matches takes up an important part
of the players' strategies.
We also point out that the dilemma captures an aspect of our
society as a series of competitions,
which is an essential part of our daily life from television shows like
Project Runway to presidential elections with the two-round system.
A sports tournament serves as a striking metaphor here,
as clearly seen from the fact that people often talk about fair play,
front-runners and a knock-out punch in these activities as well.
In order to investigate this dilemma, we consider a simple model
for a two-round tournament among four players with equal stamina. The
players have to decide how much stamina to use
in the first round (i.e., semifinals) provided that only the rest is
available in the second round. The second rounds take place between winners
(losers) of the first round for the championship (third place). We assume that
the outcome of a match is described by a well-defined probability function.
This function should contain all the essential information of the sport under
consideration, such as the rules to decide the winner (for additional discussions about the dynamics of competitions see Refs.~\cite{BenNaim2006a,BenNaim2006b}.)
It gives the probability for a player to defeat an opponent as a function of
their invested stamina, so it will be called the winning
probability function~\cite{tournament}.
The chance is 50:50 when two players spend the same amount of
stamina and it is reasonable that the winning probability is a
nondecreasing function of the stamina difference. At the end of the
tournament, each player gets a prize (payoff) for the finishing place.
We consider a general payoff
structure under a plausible constraint that the payoff never decreases as a
player moves up in rank.

Having defined the game by choosing the winning probability function together
with the payoffs,
we have to ask ourselves what we mean by solving this game.
In game theory, the Nash equilibrium is the most well-known
solution concept: Once it is achieved,
players cannot be better off by changing their own strategy
alone~\cite{Nash}. One may also investigate the game from an
evolutionary point of view. The replicator dynamics is widely used to
study the evolution of an infinite population with pure
strategies~\cite{Smith1974, Taylor1978, Hofbauer1979, rd}. In this study, we apply both methods to our
tournament model for a better understanding and compare the results to
check their consistency.

In terms of Nash equilibrium, we find three different phases,
which divide the payoff space into three regions.
(i) First, when the champion wins a much bigger payoff than the others,
any pure strategy can constitute a Nash equilibrium
as long as all four players adopt it in common.
(ii) If the payoff for the second-place winner is also valuable enough,
one has no reason to fight in the second round. Therefore, the only Nash
equilibrium is such that everyone
spends all the stamina in the first round,
i.e, semifinals.
(iii) Finally, if the payoff for last place is much smaller than the others,
a Nash equilibrium emerges when everyone adopts a mixed strategy using all
or none of the stamina with proper weights.

The above three phases are also analytically tractable from the
viewpoint of evolutionary dynamics when
the winning probability function is the Heaviside step function of
the stamina difference between two players.
The results of the replicator dynamics
are as follows. For case (i), the population evolves to a single
species adopting a common pure strategy. Which pure strategy to adopt
in the long run depends on the initial strategy distribution.
For case (ii), the population evolves to a single species using all the
stamina in the first round.
For case (iii), the population becomes a genetic polymorphism of two
species.
One spends all the stamina in the first round,
while the other reserves it all for the second round.
The proportions of these species correspond to the weights
of the mixed-strategy Nash equilibrium in case (iii). These solutions
are fully consistent with the analysis of Nash equilibria.

This paper is organized as follows. In the next section
we define our game of the two-round tournament in detail
by specifying its strategy space and payoff structure.
In Sec.~\ref{sec:Nash} we focus on symmetric Nash equilibria of this game.
We begin with simple limiting payoff structures and then proceed to
a general payoff structure to find Nash equilibria for the entire payoff
space. This analysis is followed in Sec.~\ref{sec:ED}
by the evolutionary dynamics
of an infinite population with pure strategies and a comparison of the results
with the Nash equilibria.

\section{Model}
\label{sec:Model}
Let us consider four equally strong semifinalists
and denote them by $A$, $B$, $C$, and $D$, respectively.
In the first round, player $A$ meets $B$, while $C$ meets $D$.
The second round takes place between winners (for the
championship) and between losers (for third place) of the
first round. The players' payoffs are given according to the tournament
result. The numerical values of the payoffs are denoted by $s$ for the champion,
$u$ for the second-place winner, $v$ for the third-place winner, and $w$ for
last place, where $w\le v\le u\le s$.
Nash equilibria and replicator dynamics are
invariant under translation and rescaling of the payoffs by a positive factor,
so we may set $w=0$ and $s=1$ without loss of
generality. Then the payoff parameter space reduces to the $(u,v)$ plane
with $0\le v\le u \le 1$.

Each player's strategy determines how much stamina will be spent
in the first round provided that the rest is available in the
second round. We assume that all the players have an equal amount of
stamina at the beginning and normalize it as $1$.
Then player $i$'s strategy, which is generally a mixed one,
is expressed by a normalized distribution function $\phi_i(x)$
of the mixing weights over a closed interval $[0,1]$, i.e., $\int_0^1 \phi_i(x)
dx = 1$.
If the player adopts a pure strategy of investing $x_i$ in the first round,
for example, $\phi_i(x) = \delta(x-x_i)$.
As a slight abuse of notation,
we will often abbreviate such a pure strategy as $x_i$.

Suppose that the players have formed a strategy profile~\cite{Szabo:2007uy}
$\left(\phi_A, \phi_B, \phi_C, \phi_D \right)$.
In order to calculate expected payoffs,
we need the winning probability function $f(x_i,x_j)$.
It tells us how likely player $i$ is to defeat player $j$
when they use stamina $x_i$ and $x_j$, respectively.
We furthermore assume that $f$ is a nondecreasing function of the stamina
difference $\Delta x_{ij} \equiv x_i-x_j \in [-1, 1]$,
i.e., $f(x_i,x_j) \equiv f(x_i-x_j) = f( \Delta x_{ij} )$.
This implies
that a player's winning probability never decreases as that player's investment
increases.
We do not consider a draw as a match outcome and require $f(\Delta
x_{ij})+f(\Delta x_{ji}) = 1$.
Then we have $f(0) = \frac{1}{2}$, i.e., two players spending the same amount
of stamina have an equal chance to win the match.
It is convenient to consider only the relative difference from $f(0)$ by
defining $h(\Delta x) \equiv f(\Delta x) - f(0)$, which is an odd
function, because $h(\Delta x)+h(-\Delta x) =0$.
It is non-negative for $\Delta x\ge 0$ and has a maximum at $\Delta
x=1$. The maximum is bounded by $\frac{1}{2}$ from above, because $f(\Delta x)=
h(\Delta x) + \frac{1}{2} \le1$.

To demonstrate how to calculate the expected payoffs, let us assume that
the players' moves are $x_A, x_B, x_C$, and $x_D$, respectively.
For player $A$, the probability to win the semifinal is defined as
$\frac{1}{2}+h_{AB}$, where $h_{ij} \equiv h(x_i-x_j)$.
If player $A$ has really made it, the remaining stamina for the final
must be $1-x_A$. Then, who will be $A$'s next opponent in the final?
With probability $\frac{1}{2} + h_{CD}$,
it is player $C$ with remaining stamina $1-x_C$. Player $A$ will defeat $C$
with probability $\frac{1}{2} + h[(1-x_A)-(1-x_C)] = \frac{1}{2} - h_{AC}$.
Alternatively,
the opponent can be player $D$ with probability $\frac{1}{2} - h_{CD}$, and
player $A$ will
defeat $D$ with probability $\frac{1}{2} - h_{AD}$.
Therefore, we find the probability of $A$ to be the champion as
\begin{eqnarray}
\left( \frac{1}{2} + h_{AB} \right)
\left [\left( \frac{1}{2} + h_{CD} \right)
\left( \frac{1}{2} - h_{AC} \right) \right.\nonumber\\
\left. +
\left( \frac{1}{2} - h_{CD} \right)
\left( \frac{1}{2} - h_{AD} \right) \right].
\label{eq:win}
\end{eqnarray}
It is straightforward to find the
probability to take second or third place as well.
Then player $A$'s expected payoff $\pi_A$ for these particular moves $x_A$,
$x_B$, $x_C$, and
$x_D$ is expressed as
\begin{eqnarray}
\pi_A =
\left( \frac{1}{2} + h_{AB} \right) \left [
\left( \frac{1}{2} + h_{CD} \right)
\left( \frac{1}{2} - h_{AC} \right) \right. \nonumber\\
\left. +
\left( \frac{1}{2} - h_{CD} \right)
\left( \frac{1}{2} - h_{AD} \right) \right] \nonumber\\
+ u
\left( \frac{1}{2} + h_{AB} \right) \left [
\left( \frac{1}{2} + h_{CD} \right)
\left( \frac{1}{2} + h_{AC} \right) \right. \nonumber\\
\left. + \left( \frac{1}{2} - h_{CD} \right)
\left( \frac{1}{2} + h_{AD} \right) \right] \nonumber\\
+ v
\left( \frac{1}{2} - h_{AB} \right) \left [
\left( \frac{1}{2} - h_{CD} \right)
\left( \frac{1}{2} - h_{AC} \right) \right. \nonumber\\
\left. +
\left( \frac{1}{2} + h_{CD} \right)
\left( \frac{1}{2} - h_{AD} \right) \right].
\label{eq:pay}
\end{eqnarray}
This should be averaged over the distribution functions $\phi_A (x_A)$,
$\phi_B (x_B)$, $\phi_C (x_C)$, and $\phi_D (x_D)$ to yield the eventual
expected payoff for player $A$.
The other players' payoffs are readily obtained by permuting the indices.

Note that our model is similar to the war of attrition (WA) in that each strategy is defined on a continuous space, leading to an infinite-sized payoff matrix. In addition, the WA bears a strong similarity to the chicken game, as ours does when it matters not to be the loser. One might even think of the semifinalists' dilemma as a variant of this famous game, where each player's remaining resources after the bid become as much important as the victory itself. Obviously, such a variation would just reflect the existence of the second round in our case. In addition, this modification will induce a player to give up the first round more easily whenever the opponent appears aggressive enough. Beyond a qualitative level, however, it is hard to predict the behavior, like the ratio between investing 0\% and 100\% in the first round, by this analogy. At the same time, we point out a subtle difference of our game in that it is essentially one's own bid in the first round that determines one's remaining resources, whereas it would rather be one's first opponent's bid in the WA~\cite{rd, Smith1974, Bishop1978}.

\section{Nash equilibria}
\label{sec:Nash}

Let us first consider a simple limiting case of the payoff structure by setting
$u=v=0$.
In other words, only the champion who has won both rounds gets a prize, so
it can be regarded as a winner-take-all system.
We will show that there exists an infinite number of Nash equilibria of pure
strategies where the players coordinate their strategies by choosing the same
value. If someone spends less in the first round than this coordination,
that player becomes more likely to lose the first match. If someone spends more, on the
other hand, that player is risking the chance to win the second round.

We can argue this statement more rigorously by taking player $A$'s viewpoint.
If $A$ uses a pure strategy $x_A$ while the other three use a common pure
strategy $x^\ast$, then $h_{CD}=0$ and $h_{AB} = h_{AC} = h_{AD} = h(x_A-x^\ast)
\equiv h$.
Equation~\eqref{eq:pay} thus reduces to
\begin{equation}
\pi_A(x_A, x^\ast, x^\ast, x^\ast) = \frac{1}{4} - h^2 \le
\pi_A(x^\ast, x^\ast, x^\ast, x^\ast) = \frac{1}{4}.
\end{equation}
Therefore, the pure-strategy profile
$S^\ast \equiv ( x^\ast, x^\ast, x^\ast, x^\ast )$
is a Nash equilibrium for any $x^\ast \in [0\ 1]$ if $u=v=0$.

\subsection{Region I}

Now let us check how robust the Nash equilibrium $S^\ast$ is for a
general payoff structure. Player $A$'s expected payoff of
Eq.~\eqref{eq:pay} becomes $\pi_A(x^\ast, x^\ast, x^\ast, x^\ast) = \frac{1}{4}
(1+u+v)$ because the total prize should be equally distributed among
the four players.
If player $A$ uses $x_A$ while the other three use a common strategy $x^\ast$,
Eq.~\eqref{eq:pay} is simplified to
\begin{eqnarray}
\pi_A(x_A, x^\ast, x^\ast, x^\ast) =
\left( \frac{1}{2} + h \right)
\left( \frac{1}{2} - h \right)\nonumber\\
+ u \left( \frac{1}{2} + h \right)
\left( \frac{1}{2} + h \right) \nonumber\\
+ v \left( \frac{1}{2} - h \right)
\left( \frac{1}{2} - h \right),
\label{eq:piaxa}
\end{eqnarray}
where $h \equiv h(x_A-x^\ast)$.
For the strategy profile $S^\ast$ to be a Nash equilibrium,
$\Delta \pi_A \equiv \pi_A(x^\ast,x^\ast, x^\ast, x^\ast)-\pi_A(x_A, x^\ast,
x^\ast, x^\ast) \ge 0$ for any $x_A \in [0, 1]$.
From Eq.~\eqref{eq:piaxa} we have
\begin{eqnarray}
\Delta \pi_A &=&  h(v-u) + h^2 (1-u-v) \label{eq:delpia1}\\
&=&  h [(1-h) v - (1+h) u + h].
\label{eq:delpia2}
\end{eqnarray}
First, consider the case of $x_A > x^\ast$, i.e., $h \ge 0$.
The payoff difference $\Delta \pi_A$ is non-negative only when
\begin{equation}
v \ge \frac{1+h}{1-h} u - \frac{h}{1-h}.
\label{eq:v1}
\end{equation}
We have assumed $u \ge v$, so the inequality Eq.~\eqref{eq:v1} automatically
implies that
\begin{equation}
u \ge \frac{1+h}{1-h} u - \frac{h}{1-h},
\end{equation}
which is satisfied only when $u \le \frac{1}{2}$. If this is the case,
the right-hand side (RHS) of Eq.~\eqref{eq:v1} is a nonincreasing function of $h$
because
\begin{equation}
\frac{\partial}{\partial h} \left( \frac{1+h}{1-h} u - \frac{h}{1-h} \right) =
\frac{2u-1}{(1-h)^2} \le 0.
\label{eq:dh}
\end{equation}
Let $h_0$ be the greatest lower bound of $h(x_A-x^\ast)$ for $x_A>x^\ast$.
If a certain payoff structure represented by $(u,v)$
satisfies Eq.~\eqref{eq:v1} for $h=h_0$, it is guaranteed by Eq.~\eqref{eq:dh} that
Eq.~\eqref{eq:v1} holds true for any $h \ge h_0$.
In other words, $\Delta \pi_A$ remains non-negative for any $x_A > x^\ast$ as
long as
\begin{equation}
 v \ge \frac{1+h_0}{1-h_0}\, u - \frac{h_0}{1-h_0}.
\label{eq:v2}
\end{equation}
If $x_A < x^\ast$, on the other hand, we have $h \le 0$.
Rewriting Eq.~\eqref{eq:delpia1} as
$\Delta\pi_A = h^2 \left[ (1-u-v) + |h|^{-1} (u-v) \right]$,
we see that it is always non-negative for $v\le u\le \frac{1}{2}$.

Therefore, a strategy profile
$S^\ast = ( x^\ast, x^\ast, x^\ast, x^\ast )$
is a Nash equilibrium for any $x^\ast \in [0, 1]$
if the payoff structure $(u,v)$ satisfies the inequality Eq.~\eqref{eq:v2}.
We call this area region I in the $(u,v)$ payoff space.
This region is not observable if $h_0$ vanishes: For example,
if $\Delta x$ can be infinitesimally small and $h(\Delta x)$
continuously approaches zero as $\Delta x \rightarrow 0$, then $S^\ast$ remains
as a Nash equilibrium only for $0 \le v=u\le \frac{1}{2}$ unless $x^\ast=1$.
On the other hand, region I becomes the largest when $h_0 = \frac{1}{2}$,
i.e., when $f(\Delta x)$ is the Heaviside step function as follows:
\begin{equation}
f(\Delta x) = \left\{\begin{array}{ll}
      0            & \mbox{~~} \Delta x<0\\
      \frac{1}{2}  & \mbox{~~} \Delta x=0\\
      1            & \mbox{~~} \Delta x>0.
      \end{array}  \right.
\label{eq:step}
\end{equation}
In this case, the region is bounded by $v \ge 3u-1$ and
$0 \le v \le u \le \frac{1}{2}$.

\subsection{Region II}

\begin{figure}[htb!]
\begin{center}
\includegraphics[width=0.49\textwidth]{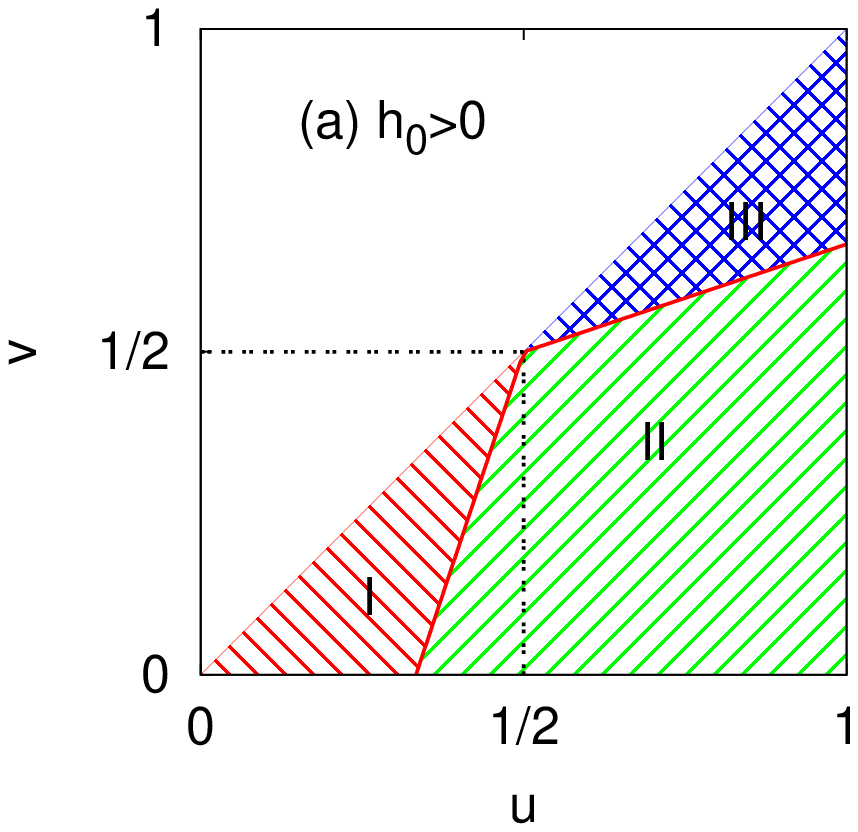}
\includegraphics[width=0.49\textwidth]{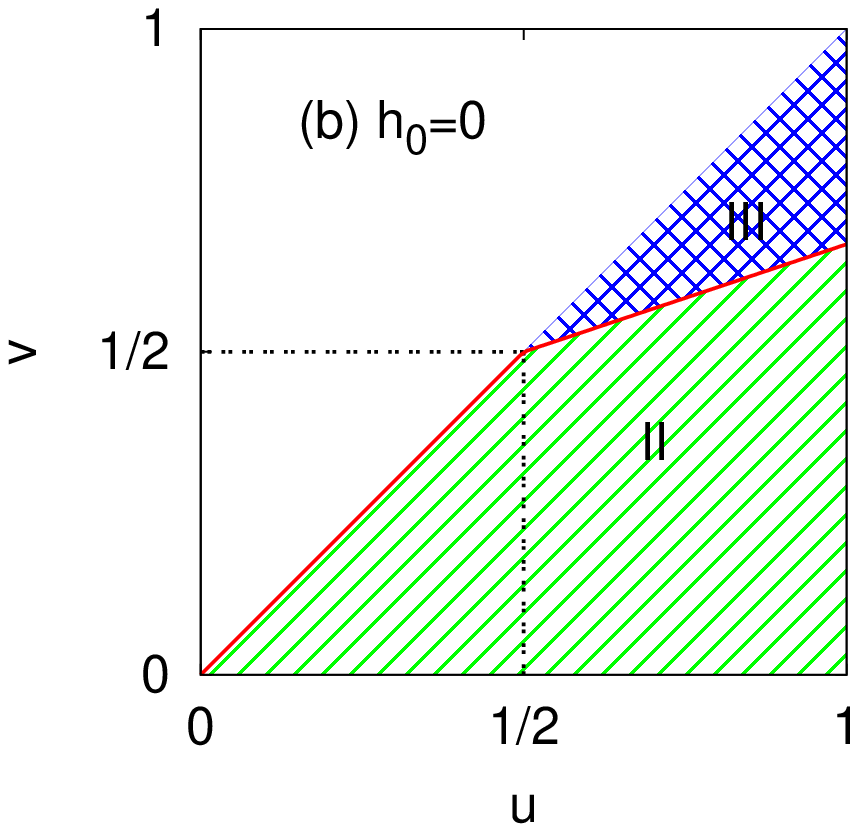}
\end{center}
\caption{(Color online)
Schematic phase diagrams with a general payoff structure, where $u$ is
the relative payoff of the second-place winner and $v$ is that of the
third-place winner, for (a) $h_0 > 0$ and (b) $h_0 = 0$.
The region of interest $u\ge v$ is
divided into three regions, denoted by I, II, and III, respectively.
Region I is characterized by the strategy profile
$S^\ast = ( x^\ast, x^\ast, x^\ast, x^\ast )$
for any pure strategy $x^\ast \in [0,1]$ and bounded by
Eq.~\eqref{eq:v2}. However, as drawn in (b),
it shrinks to a line segment of $0 \le u=v \le \frac{1}{2}$ if $\min_{\Delta
x>0} h(\Delta x) = 0$.
In region II bounded by Eq.~\eqref{eq:v4},
$S^\ast$ is not a Nash equilibrium unless $x^\ast = 1$.
If $v$ is large enough,
the players may have a common mixed strategy as a Nash
equilibrium in region III (see the text for details).
}
\label{fig:uv}
\end{figure}

If $x^\ast=1$, it is impossible to have $x_A > x^\ast$
and we only need to consider $h\le 0$.
Then, $\Delta \pi_A$ expressed in Eq.~\eqref{eq:delpia1}
is always non-negative for $u\le \frac{1}{2}$ because
both terms on the RHS are non-negative due to $v \le u$.
For $u > \frac{1}{2}$, it is better to work with Eq.~\eqref{eq:delpia2},
which tells us that $\Delta \pi_A$ is non-negative if
\begin{equation}
v \le \frac{1+h}{1-h}u - \frac{h}{1-h}.
\label{eq:v3}
\end{equation}
As shown in Eq.~\eqref{eq:dh}, if $u > \frac{1}{2}$, the RHS of Eq.~\eqref{eq:v3} only
grows as $h \equiv h(x_A - 1)$ increases.
Therefore, once $(u,v)$ satisfies Eq.~\eqref{eq:v3} for the minimum of
$h(x_A-1)$, the inequality holds true for any $x_A$.
Due to the fact that $h(\Delta x)$ is a nondecreasing
function, the minimum of $h(x_A-1)$ should be $h(-1) = -h(1) \equiv -h_1$.
We thus conclude that the strategy profile
$S_1= (1,1,1,1)$ is
the only Nash equilibrium if
\begin{equation}
v \le \frac{1-h_1}{1+h_1} u + \frac{h_1}{1+h_1}.
\label{eq:v4}
\end{equation}
Region II is the
set of $(u,v)$ points at which $S_1$ is a Nash equilibrium
but $S^\ast= (x^\ast, x^\ast, x^\ast, x^\ast )$ is not for $x^\ast<1$.
If $h_1 = \frac{1}{2}$, for example, this region is bounded by $v \le
\frac{1}{3}(u+1)$, $v \le 3u-1$, and $0 \le v \le u \le 1$.
We depict a graphical representation of regions I and II in Fig.~\ref{fig:uv}.
Note from Eqs.~\eqref{eq:v2} and \eqref{eq:v4} that the boundaries meet
at $(u,v) = \left(\frac{1}{2},\frac{1}{2} \right)$ for any $h_0$ and $h_1$
in general. It is very interesting that the phase boundaries are determined only
by both the extreme values $h_0$ and $h_1$,
and are independent of the shape of the function $h(\Delta x)$ for $0 < \Delta x <
1$.

\subsection{Region III}

The remaining region is given by
\begin{equation}
v \ge \frac{1-h_1}{1+h_1} u + \frac{h_1}{1+h_1},
\label{eq:v5}
\end{equation}
which is referred to as region III. We assume $h_1 > 0$ because $h_1 = 0$
would mean that each player's expected payoff is trivially $\frac{1}{4}
(1+u+v)$ irrespective of one's strategy.
Let us consider a simple limiting payoff structure again by setting
$u=v=1$. This is disadvantageous only for last place,
so we may call it a loser-pay-all as a rhyme for the winner-take-all system
given above.
If only the one that has lost both matches gets nothing,
the immediate goal should be to win at least one match.
The important keyword here is concentration.
It means that it is advisable to
concentrate either on the first round or on the second round.
For this reason, we look for a mixed-strategy Nash equilibrium in the form
\begin{equation}
m_p = p\delta(x) + (1-p)\delta(1-x)
\label{eq:mp}
\end{equation}
with $p\in [0, 1]$ and indeed find a strategy profile
$S_m = ( m_{p^\ast},m_{p^\ast},m_{p^\ast},m_{p^\ast} )$ as a Nash equilibrium,
where
\begin{equation}
p^\ast = \frac{1+h_1-\sqrt{1+h_1^2}}{2h_1}.
\label{eq:puv0}
\end{equation}
The proof is given in Appendix A.
It is noteworthy that when $h_1=\frac{1}{2}$, the mixing weights $p^\ast$ and
$1-p^\ast$ are $\varphi^{-2}$ and $\varphi^{-1} = 1-\varphi^{-2}$, respectively,
where $\varphi = (1+\sqrt{5})/2$ is the golden ratio.
This means that $S_m$ is a Nash equilibrium
if the weights of concentration between the first and second rounds
are in the golden ratio.
The condition that $h_1=\frac{1}{2}$ is often
reasonable because one cannot win without making an effort,
especially if the opponent is doing the best.

We can repeat the same procedure for general payoffs in region III.
The above result is generalized by the finding that
$S_m = ( m_{p^\ast},m_{p^\ast},m_{p^\ast},m_{p^\ast} )$
is a Nash equilibrium, where
\begin{equation}
p^\ast (u,v)
= \frac{kh_1 - l - \sqrt{k^2 h_1^2 + l^2 - 2k^2+ 2k}}{2k h_1}
\label{eq:psta2}
\end{equation}
with $k \equiv 1-u+v$ and $l \equiv 1-u-v$.
In Appendix B we prove that
\begin{eqnarray}
\pi_A(x,m_{p^\ast},m_{p^\ast},m_{p^\ast}) \le
\pi_A(m_{p^\ast},m_{p^\ast},m_{p^\ast},m_{p^\ast})\nonumber\\
=\frac{1}{4}(1+u+v)
\end{eqnarray}
for any $x\in [0,1]$ under the
condition that $h(\Delta x)>0$ for $\Delta x>0$.
The equality holds at both $x=0$ and $x=1$.
In other words,
when player $A$ adopts a concentration strategy
in the form of Eq.~\eqref{eq:mp}, $A$'s payoff
is predetermined as $(1+u+v)/4$, independent of the
choice of $p$. In that sense, $m_{p^\ast}$ can be considered
as a partial equalizer strategy~\cite{Press:2012iz,Hilbe:2013ts}.
One can readily check that
Eq.~\eqref{eq:psta2} reduces to Eq.~\eqref{eq:puv0} as $(u,v) \rightarrow (1,1)$.
It is also worth noting that Eq.~\eqref{eq:psta2} vanishes,
i.e., $p^\ast\rightarrow0$, so $S_m$ converges to $S_1$ as
$v$ approaches the boundary given in Eq.~\eqref{eq:v5}.

Interestingly, $S_m$ can be shown to be the only symmetric
Nash equilibrium for $u=v=1$, when the winning probability is taken as
the Heaviside step function [Eq.~(\ref{eq:step})].
Suppose that every player is using a mixed strategy $m(x)$ given by
\begin{equation}
m(x) = p \delta(x) + q\delta(1-x) + g(x),
\end{equation}
where $p$ and $q$ are probabilities of using $x=0$ and $x=1$, respectively,
and $g(x)$ is non-negative on $(0,1)$.
The conservation of total probability requires that
$\int_{0^+}^{1^-} g(x) dx = 1 - p - q \ge 0$.
Noting that player $A$'s initial payoff must be $\frac{3}{4}$ due to symmetry,
we will check whether $A$ can be better off by adopting $m_{p^{\ast}}$.
When $A$ has adopted $m_{p^{\ast}}$, whereas the others are still using $m(x)$,
player $A$'s payoff $\pi_A$ can readily be calculated, because it depends
on $p$ and $q$ but not on the shape of $g(x)$, for $m_{p^\ast}$
consisting of $x=0$ and $1$ only. Straightforward calculation leads to
\begin{equation}
\pi_A = 1 - \frac{8p - 8p^2 + 2p^3 - (1+\sqrt{5})(q^3-2q)}{4(3+\sqrt{5})},
\end{equation}
and we can show that this is always greater than $\frac{3}{4}$ unless
$m(x) = m_{p^{\ast}}$.
Therefore, any symmetric strategy profile cannot be sustained as an equilibrium
except for $S_m$.

It is also instructive to consider other strategy profiles
without such symmetry.
For example, let us choose $f(\Delta x)$ as
\begin{equation}
f(\Delta x) = \left\{
\begin{array}{ll}
1-\frac{1}{2} e^{-\Delta x /\Gamma}  & \mbox{for ~} \Delta x>0, \\
\frac{1}{2} e^{\Delta x /\Gamma} & \mbox{otherwise,}
\end{array}\right.
\label{eq:sigmoid}
\end{equation}
where $\Gamma$ is a parameter for controlling sensitivity
to the difference of the stamina~\cite{tournament}. This choice
permits a strategy profile $S_{10} = ( 1,0,1,0 )$ (and other combinations
exchanged between $A$ and $B$, or $C$ and $D$) to be a Nash equilibrium at
$(u,v)=(1,1)$, as shown in Appendix C.
It is another possible manifestation of concentration,
the keyword of region III,
in addition to the mixed-strategy profile $S_m$.
This seems to be parallel to the following asymmetric Nash equilibrium in the WA~\cite{rd, Smith1974, Bishop1978}, i.e., one player bids zero while the other does any number equal to or higher than the value of the resource in hand. We observe a similar situation in the chicken game as well because it has one symmetric mixed-strategy equilibrium plus two asymmetric pure-strategy equilibria where one player plays dove and the other plays hawk. There seems to be a class of games that permits both symmetric and symmetry-breaking solutions.

\section{Evolutionary dynamics}
\label{sec:ED}

In this section we consider replicator
dynamics for the evolution of an infinite population with pure strategies.
When $f(\Delta x)$ is chosen as Eq.~\eqref{eq:sigmoid} with
$\Gamma \rightarrow 0$,
this evolutionary framework provides an alternative derivation or interpretation
of the Nash equilibria considered in the previous section.
The condition that $\Gamma \rightarrow 0$ can be understood as
restricting our interest to a particular case of $h_0 = h_1 = \frac{1}{2}$
[see, e.g., Eq.~\eqref{eq:step}].

\subsection{Winner-take-all}
\label{ssec:WTA}

If $u=v=0$, only the final victory matters and
there exists an infinite number of Nash equilibria where the players
coordinate their strategies.
We will review this result by introducing an evolutionary process governing
an infinite population in which each individual has a
pure strategy $x$. The distribution of $x$ inside the population at time $t$
is denoted by the probability density $P(x;t)$ and assumed to be continuous at any finite $t$.
We assume that the population evolves in such a way that successful
strategies gradually increase their portions by replacing inferior ones,
as is mathematically formulated as follows:
\begin{equation}
\frac{\partial}{\partial t} P(x;t) = \left[\pi(x;t) - \left< \pi \right>\right] P(x;t),
\label{eq:rdw}
\end{equation}
where $\pi(x;t)$ means the payoff gained by strategy $x$ at time $t$ and
$\left< \pi \right> \equiv \int \pi(x) P(x) dx$ is the average payoff of the
population.
For brevity, we will suppress the dependence on $t$ henceforth.
The evolutionary process as in Eq.~\eqref{eq:rdw} is called the replicator
dynamics~\cite{Taylor1978, Hofbauer1979}. 
Even on a continuous strategy space, the replicator dynamics is well defined with no modification~\cite{Oechssler2001}.
The initial population $P(x;t=0)$ is nonzero everywhere in the unit interval
$[0, 1]$. As mentioned above, we only consider a situation where
$f(\Delta x)$ is sharp enough to be approximated as
the Heaviside step function \eqref{eq:step} for mathematical tractability.

\begin{figure}[t]
\begin{center}
\includegraphics[width=0.35\textwidth]{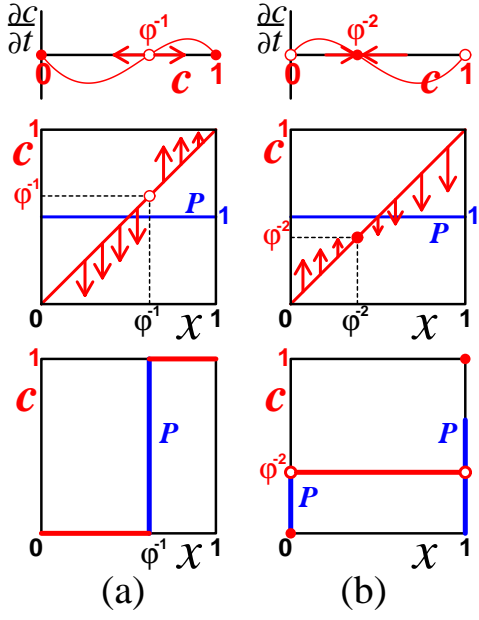}
\end{center}
\caption{(Color online)
Replicator dynamics for (a) $u=v=0$ and (b) $u=v=1$.
(a) Among the three fixed points where $\frac{\partial c}{\partial t}=0$,
the middle one
(the open circle) at $c=\varphi^{-1}$ is unstable, whereas the other two
(the closed circles) are stable. In the middle row, $c$ is shown as a function
of $x$ (the thick red line) for a uniform strategy distribution (the thin blue
line). It decreases to zero for $x<\varphi^{-1}$, while it increases to one for
$x>\varphi^{-1}$, resulting in a step function jumping at $x=\varphi^{-1}$
as shown
in the bottom row. Hence, $P(x)$ has a sharp peak at $x=\varphi^{-1}$
in the steady state. (b) The only stable fixed point is at $c=\varphi^{-2}$.
The cumulative probability $c$ decreases if it is larger than $\varphi^{-2}$
and increases when it is
smaller than $\varphi^{-2}$ resulting in the $c(x)$ configuration in the
bottom row: It is zero at $x=0$, one at $x=1$ and $c=\varphi^{-2}$ elsewhere,
regardless of the initial configuration. Hence, in the steady state,
$P(x)$ has two peaks, one at $x=0$ and the other at $x=1$.}
\label{fig:rd1}
\end{figure}

When $u=v=0$, the expected payoff for player $A$ by playing $x_A$ is equivalent
to the average of Eq.~\eqref{eq:win} over $P(x_B)$, $P(x_C)$, and $P(x_D)$, which we
will denote by $\overline{\pi}_A (x_A)$.  It is convenient to define
\begin{equation}
c(x) \equiv \int_0^1 dx' P(x') f(x-x') = \int_0^x dx' P(x'),
\label{eq:cum}
\end{equation}
and some algebra with integration by parts leads to
\begin{eqnarray}
\overline{\pi}_A (x_A)
&=& 2\int_0^1 dx_B P(x_B) f_{AB} \nonumber\\
&&\times \int_0^1 dx_C \left[
P(x_C) f_{CA} \int_0^1 dx_D P(x_D) f_{CD} \right]\nonumber\\
&=& 2 c(x_A) \int_{x_A}^1 dx P(x) c(x)\nonumber\\
&=& c(x_A) \left[ 1 - c^2 (x_A) \right],
\end{eqnarray}
where $f_{ij} \equiv f(x_i - x_j)$. Note that
$P(x) = \frac{\partial }{\partial x} c(x)$ with $c(0)=0$ and $c(1) = 1$.
One can also readily calculate the population average as
\begin{equation}
\left< \overline{\pi}_A \right> = \int_0^1 dx c(x) \left[ 1-c^2(x) \right] P(x)
= \frac{1}{4}
\end{equation}
by an integration by parts again. The fact that $\left< \overline{\pi}_A \right>
= \frac{1}{4}$ is a natural result of the symmetry among the four players,
i.e., equal probability to win on average.  Therefore, the replicator dynamics
reduces to
\begin{equation}
\frac{\partial}{\partial t} P(x) = \left[ c(x) - c^3(x) - \frac{1}{4} \right]
P(x).
\end{equation}
Equivalently, we may rewrite it as
\begin{equation}
\frac{\partial}{\partial x} \left( \frac{\partial c}{\partial t} \right) =
\frac{\partial}{\partial x} \left( \frac{1}{2} c^2 - \frac{1}{4} c^4 -
\frac{1}{4} c \right),
\end{equation}
which means that
\begin{equation}
\frac{\partial c}{\partial t}=
\frac{1}{2} c^2 - \frac{1}{4} c^4 -
\frac{1}{4} c + \xi(t),
\end{equation}
where $\xi(t)$ is a function of $t$ only. Because $c(x=0;t) = 0$ and $c(x=1;t) =
1$, $\xi(t)$ simply turns out to be $0$. Therefore, we have
\begin{equation}
\frac{\partial c}{\partial t} =
-\frac{1}{4} c (c-1) (c^2 + c - 1).
\label{eq:reduc}
\end{equation}
The RHS vanishes at $c_1=0$, $c_2=\varphi^{-1} = (\sqrt{5}-1)/2
\approx 0.618$, and $c_3=1$. Plotting
Eq.~\eqref{eq:reduc} shows that $\partial c/\partial t$ is negative between
$c_1$ and $c_2$, and positive between $c_2$ and $c_3$ [see Fig.~\ref{fig:rd1}(a)].
In other words, $c_2$ is an unstable fixed point whereas the other two are
stable.
Therefore, if we consider $x =\tilde{x}$ such that
$c(\tilde{x}) = \varphi^{-1}$, the function $c(x)$ converges to zero when
$x<\tilde{x}$, whereas it converges to one when $x>\tilde{x}$,
which means that a sharp peak in $P(x)$ develops at $\tilde{x}$ as $t$
increases.
Note that it is the cumulative probability that determines $\tilde{x}$,
so the peak position $\tilde{x}$ depends on the choice of the
initial distribution $P(x;t=0)$. In addition, Eq.~(\ref{eq:reduc}) provides
an explicit example of mapping the replicator dynamics
onto a reaction system, parametrized in terms of $c(x)$~\cite{inno,fads}.
To sum up, the replicator dynamics selects a single point $\tilde{x}$
characterized by $c(\tilde{x}) = \varphi^{-1}$ as a refined solution
among infinitely many possibilities.

\subsection{Loser-pay-all}
\label{ssec:LPA}

It is also straightforward to recast the case of $u=v=1$ into the evolutionary
framework. Let $L_A$ be player $A$'s probability to be last; we
denote its average over $P(x_B)$, $P(x_C)$, and $P(x_D)$ as
$\overline{L}_A$. The gained payoff of $A$ corresponds to the probability not to be
last; it can be written as $\overline{\pi}_A = 1-\overline{L}_A$.
One can readily check that $\overline{L}_A = (1-c)(2c-c^2)$ and
the corresponding replicator dynamics is given as
\begin{equation}
\frac{\partial}{\partial t}P(x) = - \left[ \overline{L}_A(x) - \left<
\overline{L}_A \right> \right] P(x).
\label{eq:rdl}
\end{equation}
The symmetry implies $\left< \overline{L}_A \right> = \frac{1}{4}$ again, so
Eq.~\eqref{eq:rdl} reduces to the following:
\begin{equation}
\frac{\partial c}{\partial t} = -\frac{1}{4} c(c-1)(c^2-3c+1).
\label{eq:reduc2}
\end{equation}
The RHS vanishes at $c = 0, 1$, or $\varphi^{-2} = 1 - \varphi^{-1} =
(3-\sqrt{5})/2 \approx 0.382$, werein only $c=\varphi^{-2}$ is the stable
fixed point [see Fig.~\ref{fig:rd1}(b)]. This shows that the replicator
dynamics reproduces the common mixed-strategy Nash equilibrium at $u=v=1$
on the population level.

\begin{figure}[t]
\begin{center}
\includegraphics[width=0.5\textwidth]{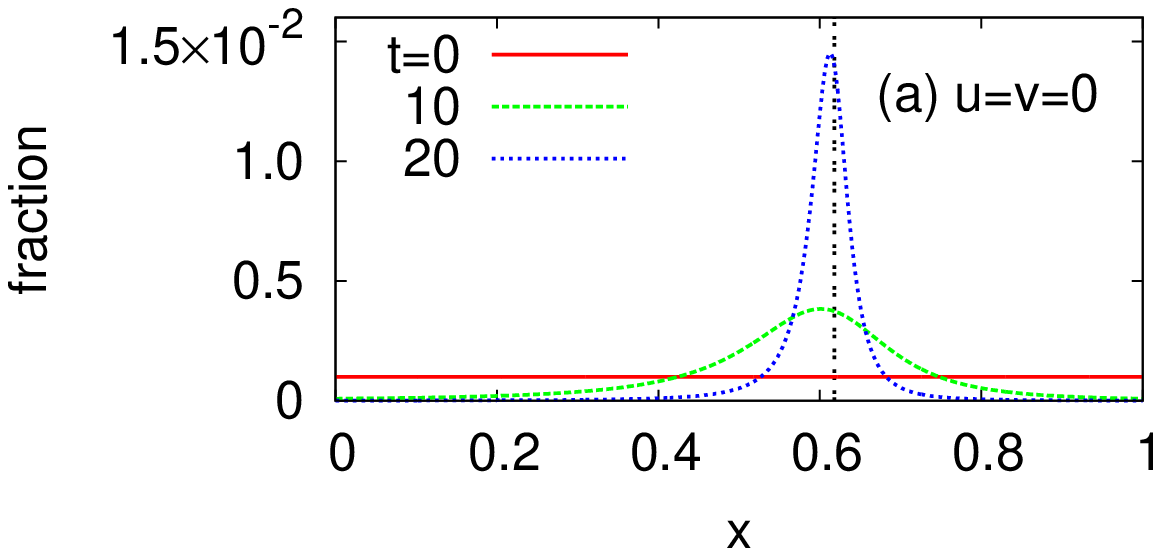}
\includegraphics[width=0.5\textwidth]{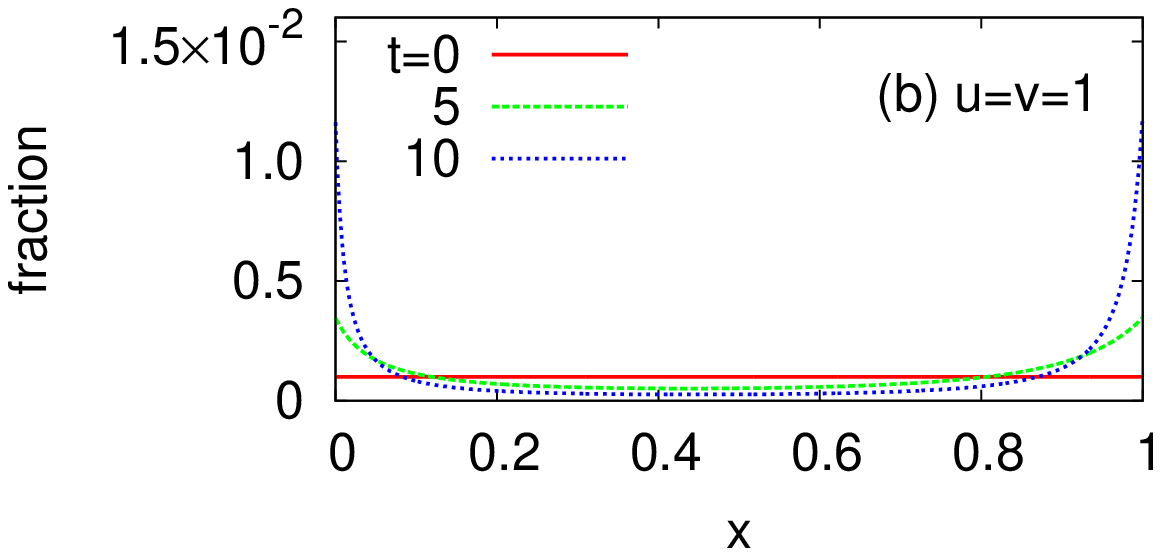}
\end{center}
\caption{(Color online) Fraction of population choosing a pure strategy $x$ at time $t$,
obtained by numerical integration of the replicator equation
with spatial and temporal step sizes $dx = 10^{-3}$ and $dt = 10^{-2}$. 
The initial distribution is assumed uniform so that each $x$
has probability $P(x;t=0) dx = 10^{-3}$.
(a) The distribution converges toward $x=\varphi^{-1}$
(the dotted vertical line)
in the winner-take-all case ($u=v=0$), while
(b) Two peaks emerge at $x=0$ with weight $\varphi^{-2}$ and $x=1$
with weight $\varphi^{-1}$ in the loser-pay-all
case ($u=v=1$). Although the simulation is a discretized version of the
replicator equation due to the nature of numerical simulation,
the results nicely confirm the analytic solutions for continuous strategy
space.
}\label{fig:rd_result}
\end{figure}

\subsection{General payoff systems}
\label{ssec:general}

For general $u$ and $v$, the same averaging process as above
yields the following equation:
\begin{equation}
\frac{\partial c}{\partial t} = \frac{1}{4}c(c-1) \Omega(c;u,v),
\end{equation}
where $\Omega(c;u,v) \equiv (u-v-1)c^2 + (u+3v-1)c + (u-3v+1)$.
The main questions are the location of $c^\ast$ such that $\Omega(c^\ast;u,v)=0$
and its stability. From $1-u+v \neq 0$, it is straightforward to find
\begin{equation}
c^\ast_\pm (u,v) = \frac{-(1-u-3v) \pm \sqrt{D(u,v)}}{2(1-u+v)},
\label{eq:cast}
\end{equation}
with $D(u,v) \equiv (1-u-3v)^2 + 4(1-u+v)(u-3v+1)$.
Note that $c^\ast_+ \rightarrow \varphi^{-1}$ as $(u,v) \rightarrow (0,0)$,
whereas $c^\ast_- \rightarrow \varphi^{-2}$ as $(u,v) \rightarrow (1,1)$.
It is worth mentioning that $c^\ast_+=1$ if $v=3u-1$,
while $c^\ast_-=0$ if $u=3v -1$, because these two lines
determine the shape of the phase diagram as depicted in Fig.~\ref{fig:uv}(a).
It is interesting that the phase diagram suggests duality
under reflection across $u+v=1$, and the replicator equation is actually
covariant under the transformation $(u,v,c,t) \rightarrow
(1-v,1-u,1-c,-t)$.

If $v > 3u-1$, which corresponds to region I in Fig.~\ref{fig:uv}(a),
only $c^\ast_+ (u,v)$ in Eq.~\eqref{eq:cast} lies inside the unit interval $[0, 1]$ as an unstable fixed
point. On the other hand, in region III where $v > \frac{1}{3}(u+1)$, only
$c^\ast_- (u,v)$ is found inside the interval and it is stable. In the
rest of the possible region of $u \ge v$, neither of $c^\ast_\pm$ is feasible,
and $c=0$ appears as a stable fixed point. This implies that $P(x)$ evolves to
$\delta(1-x)$ in the long run if the second-place prize is worth enough, which
means that all players do their best in the semifinal to advance to the final
in region II.

Furthermore, let us check the population average of the strategies in the
long-time limit, written as $\left< x \right> = \int dx \left[ x P(x;
t\rightarrow \infty) \right]$. It is simply $1$ over region II, where the
distribution is driven to a $\delta$ peak at $x=1$. In region III, the average is
found to be $1- c^\ast_-$, which is larger than or equal to $\varphi^{-1}$,
where the equality holds at $(u,v)=(1,1)$. In region I, $\left< x \right>$ is
dependent on the initial condition in general. However, if we start from a
uniform distribution $P(x,t=0)=1$, we see that $\left< x \right> = c^\ast_+$,
which is again greater than or equal to $\varphi^{-1}$,
and the equality holds at $(u,v)=(0,0)$.

\subsection{Numerical calculation}

\begin{figure}
\includegraphics[width=0.45\textwidth]{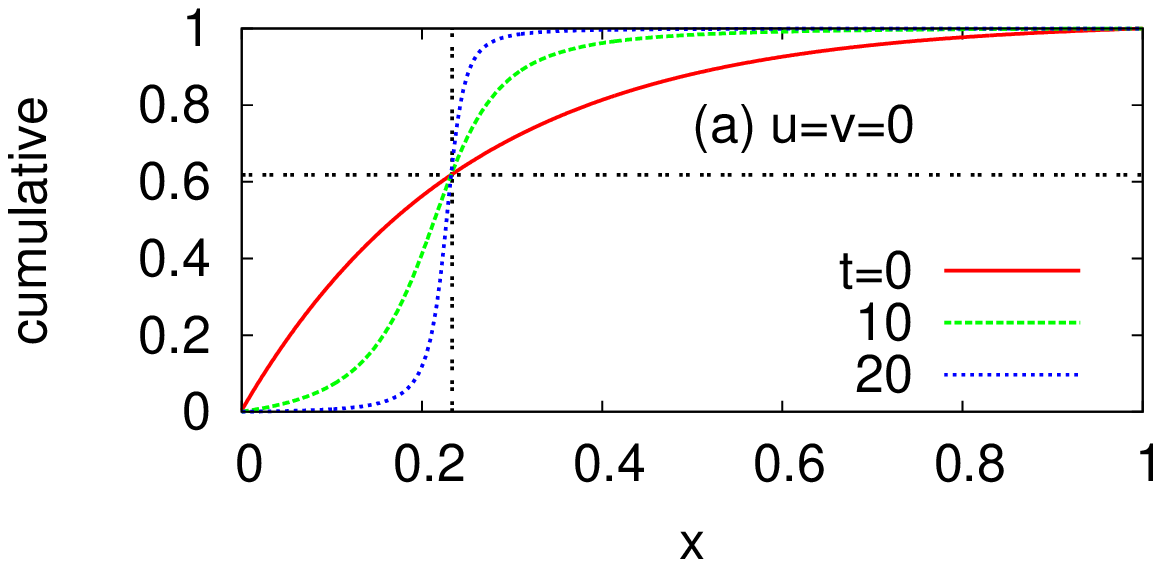}
\includegraphics[width=0.45\textwidth]{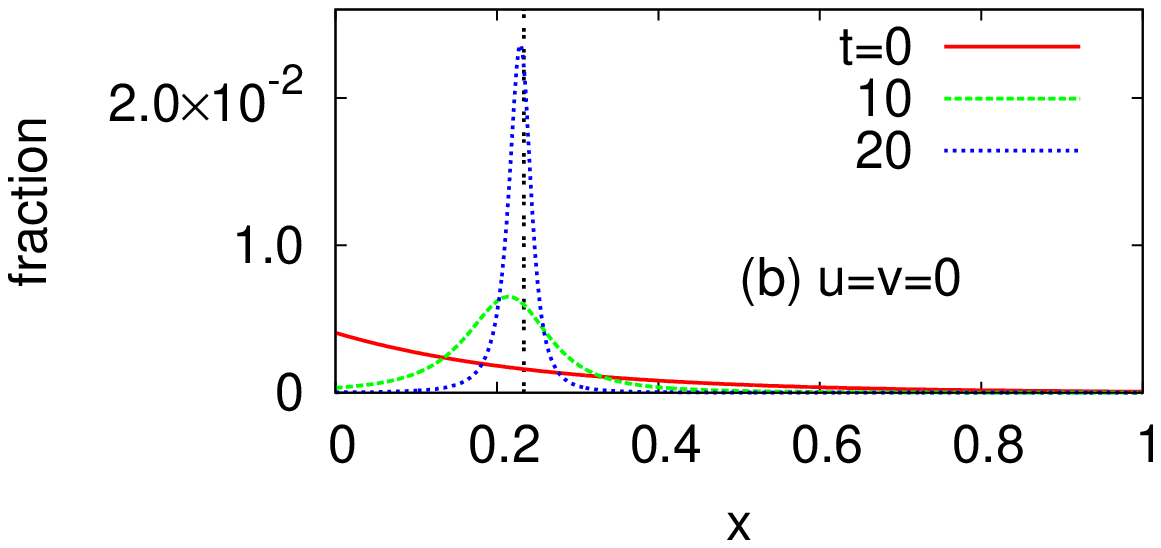}
\caption{
(Color online) (a) Cumulative $c$ and (b) the corresponding
fraction $P$ as a function of $x$ at three different values of $t$ (with
$u=v=0$), provided
that we start from an initial distribution $P(x;t=0) = e^{-4x}/Z$ with
a normalization constant $Z \equiv (1-e^{-4})/4$.
The curves are obtained by numerical integration of replicator dynamics in the
same way as explained in Fig.~\ref{fig:rd_result}.
The vertical dotted lines represent $x = \tilde{x} \approx 0.233$, and
the horizontal one does $c = \varphi^{-1}$ (see the text).
}
\label{fig:ref}
\end{figure}

To provide an intuitive example, we perform a numerical simulation of the
replicator dynamics for two representative cases: the winner-take-all ($u=v=0$)
and loser-pay-all ($u=v=1$) cases with $f(\Delta x)$ given by
Eq.~(\ref{eq:step}).
Figure 3 shows the fraction inside the population choosing a pure strategy $x$
at time $t$. We assume the initial distribution to be uniform such as $P(x;t=0) dx
= 10^{-3}$ because we assume a finite resolution $dx=10^{-3}$ for the simulation.
The population converges to $x=\varphi^{-1}$ in the winner-take-all case.
On the other hand, the loser-pay-all case shows two peaks, one at $x=0$ and
the other at $x=1$.

Another important piece of information is how quickly the Nash
equilibrium is approached in this dynamics. To answer this question,
let us linearize Eq.~(\ref{eq:reduc}) at each fixed point as follows:
\begin{equation}
\frac{\partial c}{\partial t} \approx
\left\{
\begin{array}{lcl}
-(c-c_1)/\tau_s & \text{at} & c_1=0\\
+(c-c_2)/\tau_u & \text{at} & c_2=(\sqrt{5}-1)/2\\
-(c-c_3)/\tau_s & \text{at} & c_3=1,
\end{array}
\right.
\end{equation}
where $\tau_s=4$ is a time scale to approach one of the stable fixed points
$c_1$ and $c_3$, and $\tau_u = 4/(5 - 2\sqrt{5}) \approx 7.578$
is another time scale to get away from the unstable fixed point $c_2$.
The stable fixed points lie at a distance of $O(e^{-1})$ from $c_2$, so
the overall time scale for convergence can be estimated as
$\tau \sim \tau_s + \tau_u \sim O(10)$,
which is consistent with Fig.~\ref{fig:rd_result}.
This argument gives the same $\tau$
for the loser-pay-all case when it applies to Eq.~(\ref{eq:reduc2}).

Suppose that we instead take a nonuniform distribution $P(x;t=0) =
e^{-4x} / Z$ with a normalization constant $Z \equiv (1-e^{-4})/4$ as the
initial condition. Based on the discussion in the last paragraph of
Sec.~\ref{ssec:WTA}, 
we find that the cumulative fraction $c$ equals $\varphi^{-1}$ at
$\tilde{x} =
\ln \{2e^4/[(3-\sqrt{5})e^4 + \sqrt{5}-1)]\}/4 \approx 0.233$.
This point remains invariant as an unstable fixed
point when $t$ increases, whereas $c=0$ and $c=1$ are stable
[Fig.~\ref{fig:ref}(a)]. As a consequence, $P(x)$ develops a peak at
$\tilde{x}$ as shown in Fig.~\ref{fig:ref}(b).
One can readily generalize this result to an arbitrary initial distribution
whose support is the unit interval $[0,1]$.

\begin{figure}
\includegraphics[width=0.45\textwidth]{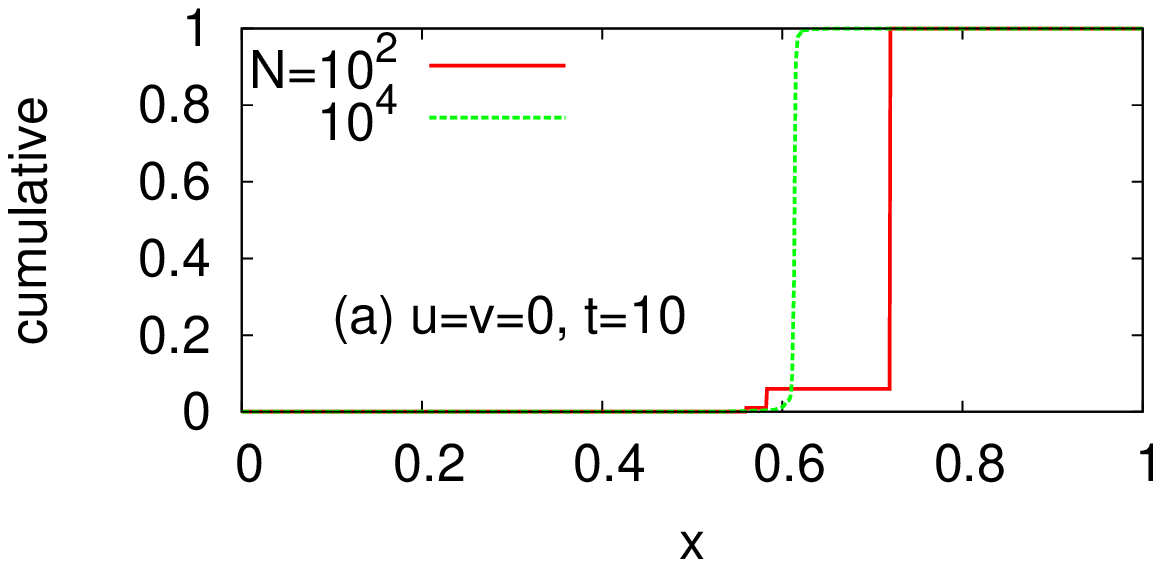}
\includegraphics[width=0.45\textwidth]{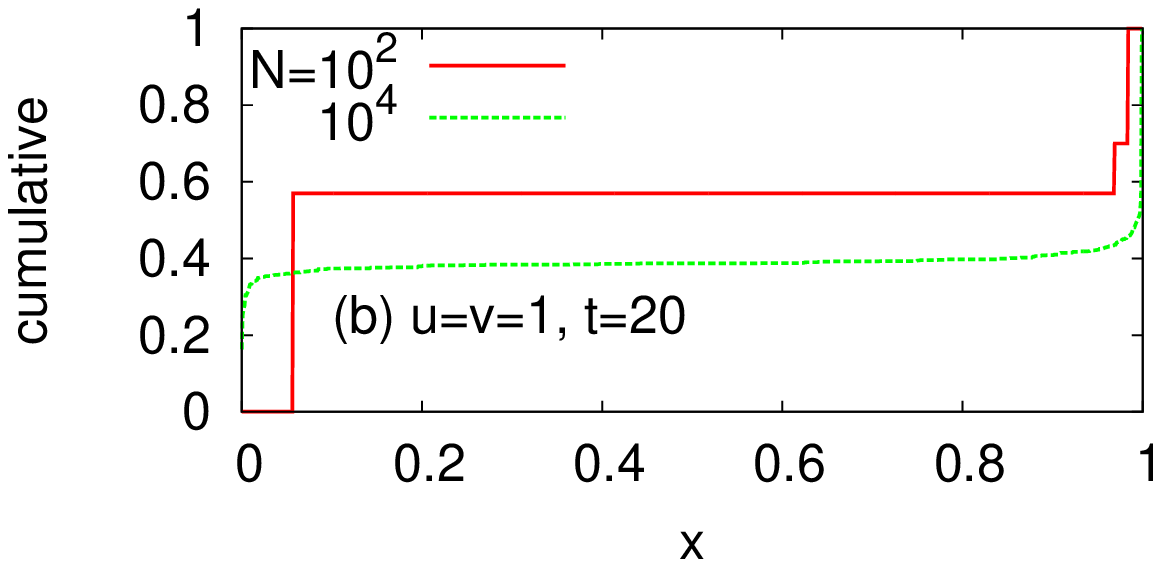}
\caption{
(Color online)
Cumulative distribution resulting from the Moran process
for a finite population of $N$ individuals. We consider $N=10^2$ and $10^4$,
and assume finite resolution of $dx=10^{-3}$
in defining strategy $x$. Each curve represents a single run
with an initial condition sampled from a uniform random distribution of $x$.
(a) In the winner-take-all case ($u=v=0$ and $t=10$),
the cumulative distribution develops a jump near $x \approx 0.6$ for $N=10^4$,
implying a single peak in the strategy distribution around $x \approx 0.6$,
after $O(10)$ time steps.
(b) For $u=v=1$, $N=10^4$, and $t=20$, the behavior is again consistent with the
result of replicator dynamics in that we find a plateau of height
$\approx 0.4$, when $t \sim O(10)$ time steps.
}
\label{fig:moran}
\end{figure}

We may also consider a finite population consisting of $N$ individuals
evolving with the Moran process (see, e.g., Ref.~\onlinecite{hcj} for a review).
In this process, we choose an individual for reproduction with probability
proportional to the payoff, which is readily calculated from
the distribution of strategies.
We then randomly choose an individual for death with equal probability,
regardless of the payoff,
and it can be the same individual that we have chosen for reproduction.
The former individual makes a copy to
replace the latter individual chosen for death and such an update is repeated
$N$ times during a single time step to give an equal chance to everyone.
For computational convenience, we assume that a strategy has
finite resolution $dx = 10^{-3}$.
Figure~\ref{fig:moran} shows typical results when initial strategies are
sampled from a uniform random distribution. When the population size is large
enough, we observe convergence with a time scale of $t \sim O(10)$
time steps for both the winner-take-all and the loser-pay-all cases.
The behavior of a small population is more complicated by the discreteness
and deserves a systematic investigation in a future study.

\section{Summary and Discussion}
\label{sec:conclusion}

We have proposed a simple example of distributing a finite amount of
resources in competitions. Despite the complexity of the decision problem among
four players, we have found two important keywords, i.e.,
balance and concentration, which can serve as a practical
guide to strategic thinking. There are two representative cases for these keywords, i.e., $u=v=0$ and $u=v=1$.
The former, in particular, has infinitely many Nash
equilibria, which are characterized by any common strategy for all the
semifinalists. The coordination would be achievable in the presence of cheap
talk, as in the case of a usual coordination game~\cite{cheap}.
In the latter, the situation is similar to the chicken game,
because it is better to give in if your opponent really goes for broke to win
the semifinal, as illustrated by $S_{10}$.
The mixed-strategy Nash equilibrium is such that you
stake all at the semifinals with probability $\varphi^{-1} \approx
0.618$ and give up the game with $\varphi^{-2} = 1-\varphi^{-1} \approx 0.382$.
The emergence of the golden ratio is fascinating,
because it is the most important
ratio of division in mathematics, also known as the most irrational number due
to its slowest convergence in the continued fraction~\cite{frac}.
We have also shown that the mixed-strategy Nash equilibrium
is the only symmetric solution when we use the Heaviside step function to
determine the probability of winning.

We have also investigated the problem from an evolutionary point of view
by introducing the replicator dynamics. 
As mentioned above, a number of different Nash equilibria exist for some
payoff structures. The replicator
dynamics selects one of them depending on the initial population distribution.
Those Nash equilibria can be accessed via a certain learning process, i.e.,
updating the strategy based on its performance.
For the winner-take-all case, the replicator dynamics converges to a pure
strategy characterized by the golden ratio again, i.e., at position $\tilde{x}$
where the cumulative probability is equal to $\varphi^{-1}$.
On the other hand, for the loser-pay-all case, two peaks emerge, one at $x=0$
with weight $\varphi^{-2}$ and the other at $x=1$ with weight $\varphi^{-1}$.
The replicator dynamics result not only provides an alternative
derivation for the Nash equilibria on the population
level, but also sheds light on the duality behind the $(u,v)$ phase diagram in
the limit of small $\Gamma$.

If we take the Sochi 2014 Olympics as an example of $(u,v)$, the relative
price of the raw material to produce a silver medal compared with that of a gold
medal roughly corresponds to $u \approx 0.6$ and that of a bronze medal amounts
to $v < 1\%$. This parameter set belongs to region II, where
everyone cares only about the semifinals in the long term.  However, this only
proves that the raw material price is a poor measure of assessing the true
values of the medals, because the finals in the Olympics have remained thrilling
throughout the century.

From a little different viewpoint, our work suggests how to design an incentive
system to affect the behavior of individuals under structured competition:
We can imagine members of an organization who compete to win a position in the
hierarchy with a limited amount of resources.
If only the top position is
rewarded in effect, for example, it will signal to the members that the
organization favors generalists rather than specialists, making them
conservative in investing effort into specific tasks.
They will even experience a social dilemma when the induced behavioral
characteristics contradict the organization's goals.
As mentioned above, our life is shaped to a great extent
by a series of competitions in an organized society.
In this respect, our tournament model will serve as a
starting point to investigate the effects of structured competition
on our behavior in various contexts.

\begin{acknowledgments} 
This work was supported by the Pukyong National University Research
Fund through Grant No. C-D-2013-1335 (S.K.B.) and the National
Research Foundation of Korea Grant funded by the Ministry of Science,
ICT and Future Planning through Grant No. 2014R1A1A2057396 (S.-W.S.) and
by Ministry of Education, Science and Technology (MEST)
through Grant No. NRF-2010-0022474 (H.-C.J.).
\end{acknowledgments}

\appendix
\section{Nash equilibrium for $u=v=1$}
\label{app:lpa}

We will prove that a strategy profile
$S_m = ( m_{p^\ast},m_{p^\ast},m_{p^\ast},m_{p^\ast} )$
is a Nash equilibrium for $u=v=1$,
where
$m_p = p\delta(x) + (1-p)\delta(1-x)$ and
$p^\ast = \frac{1+h_1-\sqrt{1+h_1^2}}{2h_1}$ with $h_1 \equiv h(1) > 0$.

Suppose that player $A$ spends $x$ in the semifinal, while the others have
a certain mixed strategy $m_p$.
We define $L_A(x,m_p,m_p,m_p)$ as the probability that $A$ becomes last in the
tournament in this situation.
It is a product of two probabilities $L_1$ and $L_2$: The former is the
probability to lose the semifinal with expending $x$. The latter is a
conditional probability to lose the third-place playoff
given that $A$ has already lost the first match with expending $x$.

Let us begin with the semifinal between $A$ and $B$. Player $B$ spends all
or nothing with probability $1-p$ and $p$, so $B$'s probability to defeat $A$
is given as
\begin{equation}
L_1 = p f(-x) + (1-p) f(1-x).
\end{equation}
For the third-place playoff, player $A$ has remaining stamina $1-x$. To
calculate $L_2$, however, we need to know $A$'s opponent's characteristics,
resulting from the other semifinal between $C$ and $D$.
For the semifinal between $C$ and $D$, there are three possibilities
in their strategic choices:
(i) With probability $p^2$, both use the strategy of $x=0$;
(ii) with probability $(1-p)^2$, both use $x=1$;
(iii) with probability $2p(1-p)$, one uses $x=0$ and the other uses $x=1$.
For the case (iii), the probability that the one with $x=1$ to lose the
semifinal is $f(-1)$. Therefore, the probability that the loser
of the semifinal between $C$ and $D$ has used up its total stamina is
$(1-p)^2 + 2p(1-p) f(-1) \equiv q$.
In other words, this is the probability for
player $A$ to meet an opponent with no remaining stamina.
The idea is that player $A$ effectively
experiences its opponent's strategy as $m_q$. That is,
\begin{equation}
L_2 = q f(x-1) + (1-q) f(x),
\end{equation}
whereby we obtain $L_A(x,m_p,m_p,m_p)$.
Some algebra shows that $L_A(0,m_p,m_p,m_p) = L_A(1,,m_p,m_p,m_p) =
\frac{1}{4}$ when $p = p^\ast$.
For $0<x<1$, we plug the explicit expression of $p^\ast$ into
$L_A(x,m_{p^\ast},m_{p^\ast},m_{p^\ast})$ and find that
\begin{eqnarray}
&4 h_1^2 L_A (x,m_{p^\ast},m_{p^\ast},m_{p^\ast})
= [ h_1 +  (a+b) G - (a-b) h_1]\nonumber\\
&\times [h_1 + (a + b) G  + (a-b) h_1 ],
\end{eqnarray}
where $G \equiv \sqrt{1+h_1^2}-1$, $a \equiv h(1-x)$, and $b \equiv h(x)$.
Noting that $G^2 = h_1^2-2G$, we expand the RHS as
\begin{eqnarray}
&&[h_1 + (a+b) G]^2 - [(a-b) h_1]^2\\
&=& h_1^2 + 2(a+b)h_1 G + (a+b)^2 G^2 -  (a-b)^2 h_1^2\\
&=& h_1^2 + 4 ab h_1^2  + 2 (a+b)(h_1-a-b) G.
\end{eqnarray}
This expression is invariant under the exchange between $a$ and $b$,
which allows us to assume that $a \ge b$ without loss of generality.
As $h_1 \ge a = h(1-x)$, putting $a$ in place of $h_1$ results in the
following inequality:
\begin{equation}
4 h_1^2 L_A(x,m_{p^\ast},m_{p^\ast},m_{p^\ast}) \ge h_1^2+4 ab h_1^2  - 2(a+b)b G.
\end{equation}
We now replace $b$ by $a$ in the last set of parentheses to get
\begin{eqnarray}
4 h_1^2 L_A(x,m_{p^\ast},m_{p^\ast},m_{p^\ast}) &\ge& h_1^2+4 ab h_1^2
- 4 ab G\nonumber\\
&=& h_1^2 + 4ab(h_1^2 - G).
\end{eqnarray}
Finally, we use the fact that $h_1^2 - G = 1+h_1^2 - \sqrt{1+h_1^2} > 0$ to
derive
\begin{equation}
4 h_1^2 L_A(x,m_{p^\ast},m_{p^\ast},m_{p^\ast}) > h_1^2,
\end{equation}
which implies $L_A(x,m_{p^\ast},m_{p^\ast},m_{p^\ast})
> \frac{1}{4}$.

Due to the symmetry among the players, it is obvious that
$L_A (m_{p^\ast},m_{p^\ast},m_{p^\ast},m_{p^\ast}) = \frac{1}{4}$.
Because $L_A(x,m_{p^\ast},m_{p^\ast},m_{p^\ast}) >
L_A(m_{p^\ast},m_{p^\ast},m_{p^\ast},m_{p^\ast})$ for any $x \in (0,1)$,
player $A$ has no reason to choose such $x$. This argument tells us that
$A$'s best choice must be a certain
mixed strategy $m_p = p\delta(x) + (1-p)\delta(1-x)$.
However, we have already seen
that this links $A$'s payoff to $\frac{1}{4}$ irrespective of $p$,
because $L_A(0,m_{p^\ast},m_{p^\ast},m_{p^\ast}) =
L_A(1,m_{p^\ast},m_{p^\ast},m_{p^\ast}) = \frac{1}{4}$.
For this reason, player $A$ may adopt $m_{p^\ast}$ as well and
we conclude that the strategy profile
$S_m = ( m_{p^\ast},m_{p^\ast},m_{p^\ast},m_{p^\ast} )$ is a
Nash equilibrium.

\section{Mixed-strategy Nash equilibrium for general $u$ and $v$}
\label{app:uv}

We will extend the above conclusion to the interior of region III by proving
that $S_m$ with $p^\ast (u,v)$ in Eq.~\eqref{eq:psta2} is a Nash equilibrium.
It is convenient to define $k \equiv 1-u+v$, $l \equiv 1-u-v$, and
$z \equiv \sqrt{k^2 h_1^2 + l^2 - 2k^2 + 2k}$.
As above, we also have $h_1 \equiv h(1)$ which is assumed to be
$0 < h_1 \le \frac{1}{2}$.
The interior of region III is described by $k+l>0$ and $1+lh_1 < k <
1$ [compare with Eq.~\eqref{eq:v5}]. Note also that $k$ is strictly positive.
In terms of these variables, Eq.~\eqref{eq:psta2} can be rewritten as
\begin{equation}
p^\ast (u,v) = \frac{1}{2} - \frac{l + z}{2kh_1}.
\end{equation}

The outline of the proof is similar to the above one for $u=v=1$:
Suppose that player $A$
spends $x$ at the semifinal, while all the others use $m_p$.
The probability for $A$ to defeat $B$ is
\begin{equation}
W_1 = pf(x) + (1-p)f(x-1).
\end{equation}
Regarding the final, we first consider strategies in the
semifinal between $C$ and $D$ and calculate the probability $r$ that
$A$'s opponent in the final has no remaining stamina.
It is given as
\begin{equation}
r = 2p(1-p)f(1) + (1-p)^2.
\end{equation}
Therefore, the probability for $A$ to win the final is thus
\begin{equation}
W_2 = r f(1-x) + (1-r) f(-x).
\end{equation}
It is straightforward to write down $A$'s payoff as
\begin{equation}
\pi_A (x,m_p,m_p,m_p) = W_1 W_2 + uW_1 (1-W_2) + vL_1 (1-L_2),
\label{eq:piauv}
\end{equation}
where $L_1$ and $L_2$ are as defined above. Due to symmetry, we see that
\begin{equation}
\pi_A (m_p,m_p,m_p,m_p) = \frac{1}{4} (1+u+v) = \frac{1}{4} (2-l).
\end{equation}
If we require $\pi_A (0,m_p,m_p,m_p) = \pi_A (1,m_p,m_p,m_p) = \frac{1}{4}
(1+u+v)$, it is satisfied only at $p^\ast (u,v)$.

Now we have to show that
\begin{equation}
\pi_A (x,m_{p^\ast},m_{p^\ast},m_{p^\ast}) < \frac{1}{4} (1+u+v)
\end{equation}
for $x \in (0,1)$, from which it follows that $m_{p^\ast}$ constitutes a
Nash equilibrium.
We obtain an explicit expression of the left-hand side
by inserting $p^\ast (u,v)$
into Eq.~\eqref{eq:piauv}, which is written as
\begin{eqnarray}
&\pi_A (x,m_{p^\ast},m_{p^\ast},m_{p^\ast}) =
(4 a^2 h_1^2)^{-1}
\left\{ (2-b) a^2 h_1^2  \right. \nonumber \\
& + 2 \omega_1 h(1-x) [h(1-x) - h_1] \label{eq:w1} \\
& + 2 \omega_2 h(x) [h(x) - h_1] \label{eq:w2} \\
& \left. + 4 \omega_3 h(x) h(1-x) \label{eq:w3} \right\},
\end{eqnarray}
where
\begin{eqnarray}
&&\omega_1 \equiv l^2(l+z) + (2l+z)k(1-k) + k^2(1-k) h_1 \nonumber \\
&&\omega_2 \equiv l^2(l+z) + (2l+z)k(1-k) - k^2(1-k) h_1 \nonumber \\
&&\omega_3 \equiv l^2(l+z) + (2l+z)k(1-k) + k^2 l h_1^2.
\end{eqnarray}
In region III, $\omega_1$ and $\omega_2$ are positive, whereas $\omega_3$ is
negative. Using $h_1 \ge h(1-x)$,
we replace $h(1-x)$ in the first set of square brackets of Eq.~\eqref{eq:w1} by $h_1$.
Likewise, because $h_1 \ge h(x)$, we do the same with $h(x)$ in the second
set of square brackets in
Eq.~\eqref{eq:w2}.
The result is the following inequality:
\begin{eqnarray}
\pi_A (&x&,m_{p^\ast},m_{p^\ast},m_{p^\ast}) \le  \\
&&(4 k^2 h_1^2)^{-1} [(2-l) k^2 h_1^2 + 4 \omega_3 h(x) h(1-x)]. \nonumber
\label{eq:sqb}
\end{eqnarray}
Clearly, the last term in the square brackets of Eq.~\eqref{eq:sqb}
is negative as long as $h(x)>0$ for $x>0$,
which leads us to the conclusion that
\begin{eqnarray}
\pi_A (x,m_{p^\ast},m_{p^\ast},m_{p^\ast}) &\le&
(4 k^2 h_1^2)^{-1} (2-l) k^2 h_1^2 \nonumber \\
&=& \frac{1}{4} (2-l).\nonumber
\end{eqnarray}

\section{Strategy profile $S_{10}=(1,0,1,0)$ at $(u,v)=(1,1)$}
\label{app:sig}

Let us choose the following sigmoid function:
\begin{equation}
f(\Delta x) = \left\{
\begin{array}{ll}
1-\frac{1}{2} e^{-\Delta x / \Gamma}  & \mbox{for ~} \Delta x>0, \\
\frac{1}{2} e^{\Delta x / \Gamma} & \mbox{otherwise,}
\end{array}\right.
\end{equation}
where $\Gamma$ is a positive parameter to control the width.
It will be shown
that $S_{10} = (1,0,1,0)$ a Nash equilibrium with this specific choice.
Of course, other combinations exchanging players $A$ and $B$, or $C$ and $D$,
are equivalent. To prove that $S_{10}$ is a Nash equilibrium,
we consider the
following two strategic configurations from player $A$'s point of view.

First, suppose that the opponent player $B$ has thrown in the towel, i.e.,
the profile is given as
$(x,0,1,0)$.  We have to find the value of $x$ that minimizes the probability
$L_A(x)$ for player $A$ to be last.
For $L_A(x) = [1-f(x)] [f(-1) f(x-1) + f(1) f(x)]$, its
derivative is readily written in terms of $f(x)$ and $\frac{df}{dx}$. Note that
the function $f(x)$ has the following property under differentiation:
\begin{equation}
\frac{df(x)}{dx} \equiv f'(x) = \left\{
\begin{array}{lll}
\frac{1}{2\Gamma} e^{-x/\Gamma} &= \frac{1-f(x)}{\Gamma} & \mbox{for~} x>0, \\
\frac{1}{2\Gamma} e^{x/\Gamma} &= \frac{f(x)}{\Gamma} & \mbox{otherwise.}
\end{array}\right.
\label{eq:diff}
\end{equation}
Using this property, we find that
\begin{equation}
\frac{d L_A}{d x} = \Gamma^{-1} [1-f(x)] f(1) [1-2 f(x)].
\end{equation}
The above expression cannot be positive
because $\frac{1}{2} \le f(x) < 1$, which implies that player $A$ is motivated
to have $x_A = 1$ to minimize $L_A$. This choice is reasonable because $A$ must
win this match with strong possibility at any cost.

Second, as an opposite situation, suppose that player $B$ stakes all on a single
throw, i.e., where the profile is given as $(x,1,1,0)$.
If every other player is
concentrating all the efforts on either this round or the next one, player
$A$ has no reason to play by halves because such a strategy would always put
$A$ in a weaker position than the opponent.
This statement is mathematically expressed by
$\frac{d^2 L_A}{d x^2} \le 0$, where
$L_A(x) = [1-f(x-1)] [f(-1) f(x-1) + f(1) f(x)]$ is
the probability for player $A$ to be last.
The inequality implies that even if the first derivative $\frac{d L_A}{dx}$
vanishes, it will be a maximum of $L_A$, so the minima should
be found at the boundary points, i.e., either at $x=0$ or at $x=1$.
Direct calculation shows that
$L_A(x=0) - L_A(x=1) = \frac{1}{4} [f(1)-1] [2f(1)-1]^2 < 0$
for any possible $f(1)$ between $\frac{1}{2}$ and $1$.
In short, the minimum of $L_A$ appears at $x=0$. This means that
player $A$ should prepare for the next match avoiding all-out war this time to
reduce the probability to be last.

%
\end{document}